\documentclass[10pt,letterpaper,twocolumn]{article} 

\usepackage{ol2}
\usepackage[draft]{hyperref}
\usepackage{amsmath}

\begin{document}

\twocolumn[ 

\title{ A variational approach to Schr\"{o}dinger equation with parity-time symmetry Gaussian complex potential}


\author{Sumei Hu,$^{1,2}$ Guo Liang,$^1$Shanyong Cai,$^1$ Daquan Lu,$^1$Qi Guo,$^1$\\and Wei Hu$^{1,*}$}
\address{$^1$Laboratory of Photonic Information Technology, South China Normal University, \\ Guangzhou  510631, P. R. China}
\address{$^2$ Dept of physics, Guangdong University of Petrochemical Technology, \\ Maoming 525000, P. R. China}
\address{$^*$Corresponding author:huwei@scnu.edu.cn}

\begin{abstract}
A variational technique is established to deal with the
Schr\"{o}dinger equation with parity-time ($\mathcal{PT}$) symmetric
Gaussian complex potential. The method is extended to the linear and
self-focusing and defocusing nonlinear cases. Some unusual
properties in PT systems such as transverse power flow and PT
breaking points can be analyzed by this method. Following numerical
simulations, the analytical results are in good agreement with the
numerical results.

\end{abstract}

\ocis{190.0190, 190.4350.}

\maketitle 
] 

{\it INTRODUCTION}: 
People have paid much attention to optical properties in parity-time
($\mathcal{PT}$) symmetric system during recent years both
theoretically and
experimentally\cite{PT2007-OL2632,PT2008-PRL030402,
PT2008-PRL103904,PT2010-pra, PT2010-Nature, PT2009-PRL}. Bender {\it
et al} found that the non-Hermitian  Hamiltonian with $\mathcal{PT}$
symmetry can exhibit entirely real spectrum \cite{PT1998-prl}.In
optics, $\mathcal{PT}$ symmetric potentials can be also constructed
by a complex gain/loss refractive-index distribution into the
waveguide, which make the refractive index distribution is
$\mathcal{PT}$
symmetry\cite{PT2007-OL2632,PT2008-PRL030402,PT2008-PRL103904}. Many
unusual features stemming from $\mathcal{PT}$ symmetry have been
found, such as power oscillation \cite{PT2010-Zheng}, absorption
enhanced transmission \cite{PT2009-PRL},  nonlinear switching
structure \cite{PT2010-Sukhorukov}, and unidirectional invisibility
\cite{PT2011-Zin}.

But the properties of $\mathcal{PT}$ symmetry mentioned above are
mostly discussed by numerical simulations and experimental results.
In order to get a good understanding of the properties of
$\mathcal{PT}$ symmetry solitons, it is essential to present an
analytic solution, even an approximate one. Musslimani et al have
present closed form solutions to a certain class of nonlinear
Schrodinger equations involving $\mathcal{PT}$ symmetry
potential\cite{PT2008-JPA244019}. R.EI-Ganainy et al have developed
a formalism for coupled optical parity-time symmetric systems by
Lagranian principles\cite{PT2007-OL2632}. It would be very desirable
to obtain approximate analytical results for the general
$\mathcal{PT}$ symmetry Schr\"{o}dinger equations. For the nonlinear
Schr\"{o}dinger equations, the approximate solutions can provide a
better physical understanding of the interplay between
$\mathcal{PT}$ symmetry potential and nonlinear effects in
connection with solitons propagation.
  The variational method has been widely applied
to obtain approximated solutions for problems concerning beam
propagation within the framework of the NLSE\cite{variational-2007}.
Motivated by the previous works in which it was studied the
dissipative solitons solutions by the variational
approach\cite{variational-2006,variational-1998}. In this paper, we
will employ a variational approach based on the dissipative system
to deal with the Schr\"{o}dinger equation with PT symmetric Gaussian
complex potential, and some unusual properties in $\mathcal{PT}$
system can be analyzed by this method.

{\it MODEL}: 
We consider the (1+1)-dimensional evolution equation along the
longitudinal direction $z$ in linear or Kerr-nonlinear media with
complex PT potentials. The propagation of the beam is governed by
the Schr\"{o}dinger equation\cite{PT2007-OL2632}, i.e.
\begin{equation}\label{propap1}
i\partial_z U+\partial^2_x U+ [V(x)+iW(x)]U+\sigma|U|^2U= 0.
\end{equation}
Here $U$ is the slowly varying complex field envelop, $x$ is the
transverse coordinate. $\sigma=1$
represents the self-focusing propagation, $\sigma=-1$ represents the
self-defocusing propagation, and $\sigma=0$ represents the linear
situation. $V(x)$ and $W(x)$  are the real and the imaginary parts
of $\mathcal{PT}$ symmetry complex potentials respectively. The normalization relation is given in the reference \cite{PT2007-OL2632}. 

If the solution of stationary wave modes   for  Eq. (\ref{propap1}) has the form like
$U=\phi(x)\exp[i\theta(x)]\exp(ibz)$, where $b$ is the propagation
constant, $\phi(x)$ and $\theta(x)$ are real functions standing for
the amplitude and phase of light, then
\begin{equation}\label{realEQ}
\partial^2_x \phi + V(x)\phi+ \sigma|\phi|^2\phi
-b \phi = \phi (\partial_x \theta)^2
\end{equation}
\begin{equation}\label{imagEQ}
\partial_x(\phi^2 \partial_x \theta) + W(x) \phi^2 =0.
\end{equation}
It is noteworthy that  Eq. (\ref{imagEQ}) is a specialized form of Poynting's theorem for electromagnetic fields in dispersive media with losses\cite{Jacksonbook}, where the transverse component of Poynting vector can be written as $S=\phi^2 \partial_x\theta$ (only the transverse differential remains for stationary modes). The term $W(x) \phi^2$ represents the absorption(or gain) in the medium. Equation (\ref{imagEQ}) can be applicable to the general dissipative system. At the center of the $\mathcal{PT}$ system, we have 
\begin{equation}\label{PflowC}
(\phi^2\partial_x\theta)|_{x=0}
=\int_{0}^{\infty}W(x) \phi^2 dx=-\int_{-\infty}^{0}W(x) \phi^2 dx.
\end{equation}
That means the energy-flow at the center must equal to the total loss (or gain)
at each side. As a result,   the intensity at the center can not be zero for any stable solution and the dark solitons can not exist in any $\mathcal{PT}$ system.

{\it VARIATIONAL METHOD}:

 According to the literatures\cite{variational-2006,variational-1998}, the Lagrangian for  Eq.~(\ref{propap1}) is the sum of two terms, a conservative one $L_C$
 and a non-conservative one $L_{NC}$, where 
\begin{equation}\label{lc}
L_{C}=\frac{i}{2}(U\partial_z U^\prime -U^\prime \partial_z U ) +|\partial_x U|^2 -V|U|^2 -\sigma\frac{1}{2}|U|^4.
\end{equation}
The $\mathcal{PT}$ symmetric gaussian complex potentials are used here as
$V(x)=V_0\exp(-x^2)$ and $W(x)=W_0x \exp(-x^2)$,where $V_0$ and $W_0$ represent the depth of the real and the imaginary parts of $\mathcal{PT}$ potentials
respectively\cite{GSPT2011-pra}. We suppose the trial function as,
\begin{eqnarray}\label{ansatz}
U &=& a(z) \exp\left\{-\frac{x^2}{2w(z)^2}+i\theta_0(z){\rm
erf}[\frac{x}{w_1(z)}]\right\} \nonumber\\
& &\times \exp[iC(z)x^2+i\Phi(z)],
\end{eqnarray}
where $a(z)$, $w(z)$, $C(z)$, and $\Phi(z)$ stand for the
amplitude, width, chirp and the phase of the beam, $\theta_0(z)$,
and $w_1(z)$ stand for the amplitude and the width of wavefront
lean $\theta$. By substituting the trial solution into Eq.(\ref{lc}) and 
the averaged conservative Lagrangian can be obtained, 
\begin{equation}\label{aLagrangianC}
\begin{split}
<L_{C}> = & \int_{-\infty}^{+\infty}L_{C}dx \\
=&\frac{\sqrt{\pi}a^{2}w^{3}}{2}\frac{d C}{d
z}+\sqrt{\pi}a^{2}w\frac{d\Phi}{dz}
+\frac{\sqrt{\pi}a^{2}}{2w}\\
+&\frac{4a^2 \theta_0 ^2
w}{\sqrt{\pi}w_1\sqrt{2w^2+w_{1}^{2}}}+2\sqrt{\pi}C^{2}a^{2}w^3\\
-&\frac{\sqrt{\pi}
V_{0}a^{2}w}{\sqrt{w^{2}+1}}-\sigma\frac{\sqrt{2\pi}a^{4}w}{4}.
\end{split}
\end{equation}
The Euler-Lagrange equations for parameter $\eta$ is
\begin{equation}\label{EulerEq}
 \frac{d}{dz}\left(\frac{\partial
<L_C>}{\partial \eta_z}\right)-\frac{\partial <L_C>}{\partial
\eta}=2Re\left\{ \int_{-\infty}^{+\infty}Q\frac{\partial
U^{*}}{\partial \eta}dx \right\}
\end{equation}
in which $Q=-iW(x)U$ is the dissipative term in Eq. (\ref{propap1}).   
Following the process\cite{variational-2006,variational-1998}, substituting $\eta$ with $a$, $w$, $w_1$, $\theta_0$, $C$ and $\Phi$ into the Eq.~(\ref{EulerEq}), we arrive at the
relations between the parameters,
\begin{equation}\label{E_Phi1}
\frac{d(\sqrt{\pi}a^2 w)}{dz}=\frac{dP}{dz}=0,
\end{equation}
\begin{equation}\label{width1_is1}
w_1\equiv 1.
\end{equation}
\begin{equation}\label{theta_0}
\theta_0=\frac{\sqrt{\pi}W_{0}w^{2}}{4(w^{2}+1)}.
\end{equation}
Equations~(\ref{E_Phi1}) is a trial result which  means the power of a stationary modes is unchange during propagation. From Eqs.(\ref{width1_is1}) and (\ref{theta_0}), we find that the shape of the phase function $\theta$ is independent of the beam width.  Numerical simulations show that even for unstable solution,  the shape of the phase function  is very similar to the error function, and its amplitude ($\theta_0$) is mainly determined by $ W_{0}$. We also find that these properties are identical to the multimode solitons and other $\mathcal{PT}$ systems such as Scarff II potential.

Introducing $P_0=\sqrt{\pi}a^2w$, we arrive at the evolution equations,
\begin{eqnarray}
\frac{d w}{dz} &=&4wC,\label{evolution_w}\\
\frac{d C}{dz} &=& \mu(w)-4C^2-\frac{\sigma P_0}{2\sqrt{2\pi}w^3},\label{evolution_C}\\
\mu(w)&=& \frac{1}{w^4} -\frac{V_0}{(w^2+1)^{3/2}}
+\frac{W_0^2w^4}{2 (w^2+1)^2(2w^2+1)^{3/2}}, \nonumber \\
b&= &\frac{w^2+1}{2w^4}
-\frac{3w^2+1}{2}\mu(w)+\frac{V_0}{2(w^2+1)^{3/2}} \nonumber \\
&+&\sigma\frac{5P_0}{4 \sqrt{2\pi}w}.
\end{eqnarray}

\begin{figure}[htbp]
   \centering
   \includegraphics[width=8.0cm] {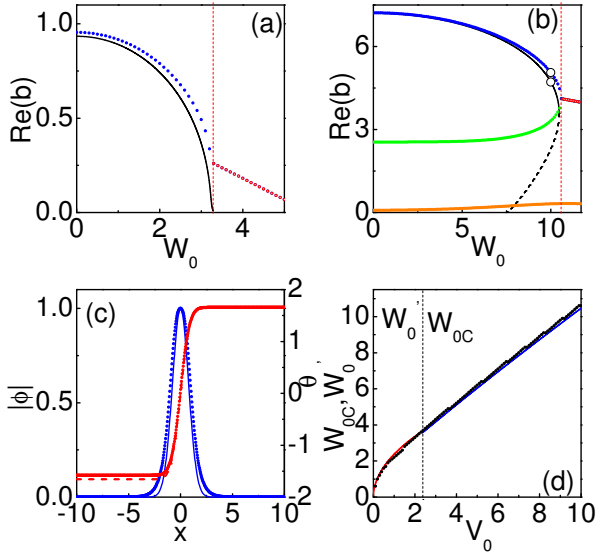}
   \caption{(color online)(a) and (b)The eigenvalues for linear $\mathcal{PT}$ modes versus the value of $W_0$ for (a) $V_0=2$ and (b) $V_0=10$.
(c) The amplitude and the nontrivial phase of the eigenfunction with
$V_0=10$ and $W_0=10$ (d)$\mathcal{PT}$-breaking point versus
$V_0$.(For all of the cases, solid: variational results,  dotted:
numerical results)}\label{linearcase}
\end{figure}

{\it LINEAR MODES}: 
For the case of stable solution, the derivatives of beam width and curvature in the evolution equations are vanished. In linear propagation,  i.e. $\sigma=0$, we can get $C=0$, $\mu(w)=0$, 
\begin{equation}\label{linearmode1}
b=\frac{w^2+1}{2w^4}+\frac{V_0}{2(w^2+1)^{3/2}},
\end{equation}
\begin{equation}\label{linearmode2}
W_0=\frac{\sqrt{2}(2w^2+1)^{3/4}}{w^4}[V_0 w^4
(w^2+1)^{1/2}-(w^2+1)^{2}]^{1/2}.
\end{equation}
From Eqs.~(\ref{linearmode2}), for a large value of $V_{0}$, we have a maximum
value of $W_{0C}$, which can be regarded as the
$\mathcal{PT}$-breaking point. For a small value of  $V_{0}$, there doesn't
exist a maximum value of $W_{0C}$, however, when $W_0^\prime
=2^{5/4}\sqrt{V_0}$, $b \rightarrow 0$, and $W_0^\prime$ can be
regarded as the $\mathcal{PT}$-breaking point. Figures
(\ref{linearcase})(a) and (b) shows the relation to the eigenvalues
with $W_{0}$ for different $V_{0}$ in numerical and variational
method. We can see that for a large $V_{0}$($V_{0}$=10), the
$W_{0C}$ obtained by the variational method is in good agreement
with the $\mathcal{PT}$-breaking point in the numerical method, and
for a small $V_{0}$ ($V_{0}=2$), the $W_0^\prime$ is close to the
$\mathcal{PT}$-breaking point. The $\mathcal{PT}$-breaking point
with different $V_{0}$ in numerical method and the relation of the
$W_{0C}$ or the $W_0^\prime$ with $V_{0}$ in analysis are shown in
the fig. (\ref{linearcase})(d). One can see that the $W_{0C}$ or the
$W_0^\prime$ is in good agreement with the $\mathcal{PT}$-breaking
point. Figure(\ref{linearcase})(c) shows that the amplitude and the
nontrivial phase of the eigenfunction are in good agreement with the
numerical results too.

{\it SOLITONS}:
For the nonlinear propagation, i.e. $\sigma = \pm 1$, we can get $C=0$ and $\sigma P_0=
2\sqrt{2\pi}w^3\mu(w)$ for the $\mathcal{PT}$ solitons. For a given
value of $V_0$ and $W_0$, the $\mathcal{PT}$ solitons obey
\begin{equation}\label{solitonmode}
\begin{split}
b=&\frac{w^2+1}{2w^4}+\frac{2w^2-1}{2}\mu(w)+\frac{V_0}{2(w^2+1)^{3/2}},\\
 P_0 = & 2 \sigma\sqrt{2\pi}w^3\mu(w).
\end{split}
\end{equation}
 Without losing of generality, we take
$V_0=10$ and $W_0=5$ in this subsection.

\begin{figure}[htbp]
  \centering
  \includegraphics[width=8.3cm] {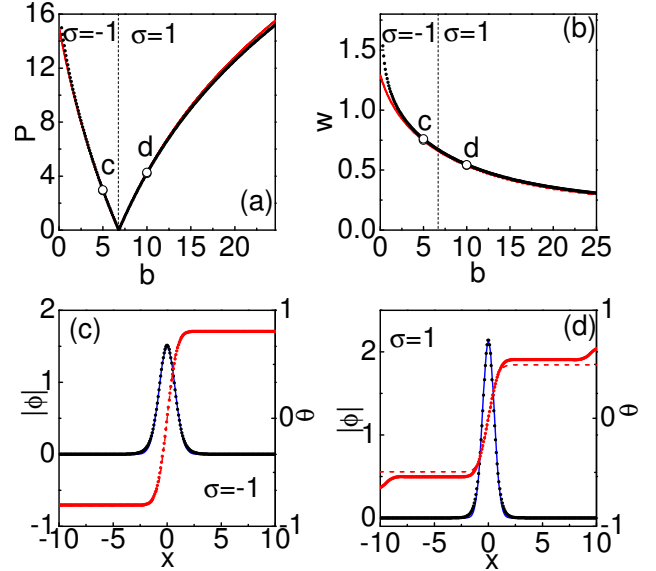}
  \caption {(Color online)(a) The  power of the solitons versus
propagation constants; (b) the beam width of the solitons versus
propagation constants;(Red: self-defocusing nonlinear, blue:
self-focusing nonlinear. );
  (c) and (d) The amplitude and the nontrivial phase of the
solitons in self-defocusing and self-focusing nonlinear
,respectively, which is corresponding to the cases marked as circle
symbols in (a) and (b). (Red: the nontrivial phase of the solitons ,
blue: the amplitude of the solitons. For all of the cases,  solid:
analytical results,  dashed: numerical results, and $V_0$=10,
$W_0$=5.
  )}\label {nonlinear}
  \end{figure}

Figures (\ref{nonlinear}) (a) and (b) are the relations between the
power and the beam width of solitons with the propagation constants,
respectively. The critical propagation constant $b_c$ at zero power
is the eigenvalue for the linear mode. When propagation constant
$b>b_c$, it is corresponding to the solitons in the self-focusing
nonlinear media, otherwise it represents the solitons in the
self-defocusing media. We can see that in the self-defocusing media,
the power of the bright solitons decreases with increasing of the
propagation constants, however, it is contrary to the bright
solitons in the self-focusing media. However, the beam width
decreases with increasing of the propagation constants in the whole
region. The amplitude and the nontrivial phase of the solitons in
numerical method and variational approach are shown in Figs.
(\ref{nonlinear})(c) and (d), which are corresponding to the cases
marked as circle symbols in Figs. \ref{nonlinear}(a) and (b). One
can see that the analytical results are in excellent agreement with
the numerical results for both self-focusing and self-defocusing
cases.

{\it Stability of the solitons}:
From the evolution equations, Eq.~(\ref{evolution}), we define
\begin{eqnarray}\label{evolution}
\begin{split}
F=\frac{d w}{dz} =&4wC,\\
G=\frac{dC}{dz} =& \mu(w)-4C^2-\frac{\sigma P_0}{2\sqrt{2\pi}w^3}
\end{split}
\end{eqnarray}
The equation associated with the  the Jacoby determinant
corresponding to the stability criterion is\cite{variational-2006}
\begin{equation}\label{Stability1}
\lambda^2+4C\lambda-32C^2-4w\frac{d\mu(w)}{dw}-\frac{6\sigma
P_0}{\sqrt{2\pi}w^3}=0
\end{equation}
When $Re(\lambda)=0$, or $\lambda^2<0$, solitons is stable.
Substituting $C=0$ and $\sigma P_0=2\sqrt{2\pi}w^3\mu(w)$ to
Eq.~(\ref{Stability1}), one has
\begin{equation}\label{Stability2}
\lambda^2=4w\frac{d\mu(w)}{dw}+12\mu(w)<0.
\end{equation}

{\it CONCLUSION}:
In conclusion, we have analyzed the properties of the beam in PT
symmetric Gaussian complex potentials by the variation method.
According to the analysis, the transverse power-flow density is
associated with the wavefront lean $\theta$ of the beam. For the
linear case, the $\mathcal{PT}$-breaking points are obtained by the
variational approach, and the eigenvalue obtained by the variational
approach are in agreement with the ones by the numerical method. For
the nonlinear case, there exists a critical propagation constant for
bright soliton existing in the self-focusing and self-defocusing
media. The relations between the power of solitons with the
propagation constants and the beam width are gotten analytically and
confirmed by numerical simulations. The amplitude and the nontrivial
phase of the solitons in both the self-focusing and self-defocusing
media are agreement with the numerical results.

{\it ACKNOWLEDGMENTS} This research was supported by the
National Natural Science Foundation of China (Grant Nos. 10804033
and 11174090), the Program for Innovative Research No. 06CXTD005),
and the Specialized Research Fund for the Doctoral Program of Higher
Education (Grant No.200805740002).

\end{document}